\documentclass[aps,prl,twocolumn,preprintnumbers,amsmath,amssymb,superscriptaddress]{revtex4-1}
\usepackage{graphicx}
\usepackage{subfigure}
\usepackage{epsfig}
\usepackage{dcolumn}
\usepackage{bm}
\usepackage{ulem}
\usepackage{color}
\usepackage{float}

\def\bea{\begin{eqnarray}}      \def\eea{\end{eqnarray}}
\def\ba{\begin{array}}
\def\ea{\end{array}}
\def\bnum{\begin{enumerate} }
\def\enum{\end{enumerate}}

\def\nn{\nonumber}

\def\PRB{Phys. Rev. B}

\def\vk{{\bf k}}
\def\Eq#1{Eq.~(\ref{#1})}
\def\Fig#1{Fig.~\ref{#1}}

\newcommand{\+}{\dagger}

\newcommand{\<}{\langle}
\renewcommand{\>}{\rangle}

\newcommand{\ra}{\rightarrow}

\newcommand{\Del}{\Delta}

\newcommand{\al}[1]{\begin{align}#1\end{align}}
\newcommand{\eq}[2]{
	\begin{equation}
	#1 \label{#2}
	\end{equation}
}
\newcommand{\md}{\mathrm{d}}
\newcommand{\re}[1]{\frac{1}{#1}}
\newcommand{\me}{\mathrm{e}}
\newcommand{\mi}{\mathrm{i}}

\renewcommand{\rm}[1]{\mathrm{#1}}

\newcommand{\vect}[1]{\boldsymbol{#1}}

\begin{document}
\title{Quantum criticality preempted by nematicity}
\author{Shi-Xin Zhang}
\affiliation{Institute for Advanced Study, Tsinghua University, Beijing 100084, China}
\author{Shao-Kai Jian}
\affiliation{Institute for Advanced Study, Tsinghua University, Beijing 100084, China}
\affiliation{Department of Physics, Harvard University, Cambridge, MA 02138, USA}
\author{Hong Yao}
\email{yaohong@tsinghua.edu.cn}
\affiliation{Institute for Advanced Study, Tsinghua University, Beijing 100084, China}
\affiliation{State Key Laboratory of Low Dimensional Quantum Physics, Tsinghua University, Beijing 100084, China}

\begin{abstract}
Exotic physics often emerges around quantum criticality in metallic systems. Here we explore the nature of topological phase transitions between 3D double-Weyl semimetals and insulators (through annihilating double-Weyl nodes with opposite chiralities) in the presence of Coulomb interactions. From renormalization-group (RG) analysis, we find a non-Fermi-liquid quantum critical point (QCP) between the double-Weyl semimetals and insulators when artificially neglecting short-range interactions. However, it is shown that this non-Fermi-liquid QCP is actually \textit{unstable} against nematic ordering when short-range interactions are correctly included in the RG analysis. In other words, the putative QCP between the semimetals and insulators is preempted by emergence of nematic phases when Coulomb interactions are present. We further discuss possible experimental relevance of the nematicity-preempted QCP to double-Weyl candidate materials HgCr$_2$Se$_4$ and SrSi$_2$.
\end{abstract}

\date{\today}
\maketitle

\begin{figure}[t]
\centering
\subfigure[]{\label{}\includegraphics[width=0.48\columnwidth]{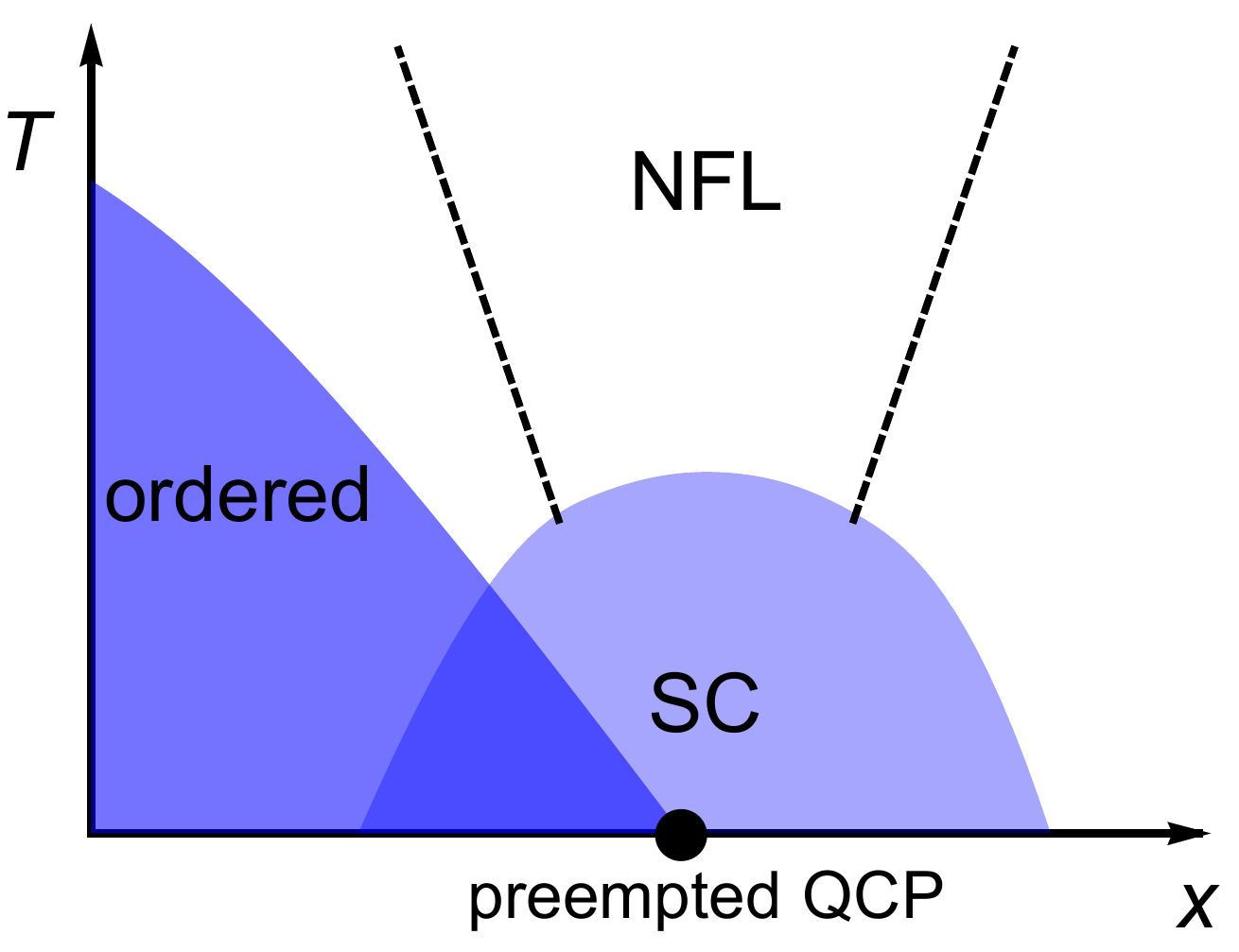}}~~
\subfigure[]{\label{IC}\includegraphics[width=0.48\columnwidth]{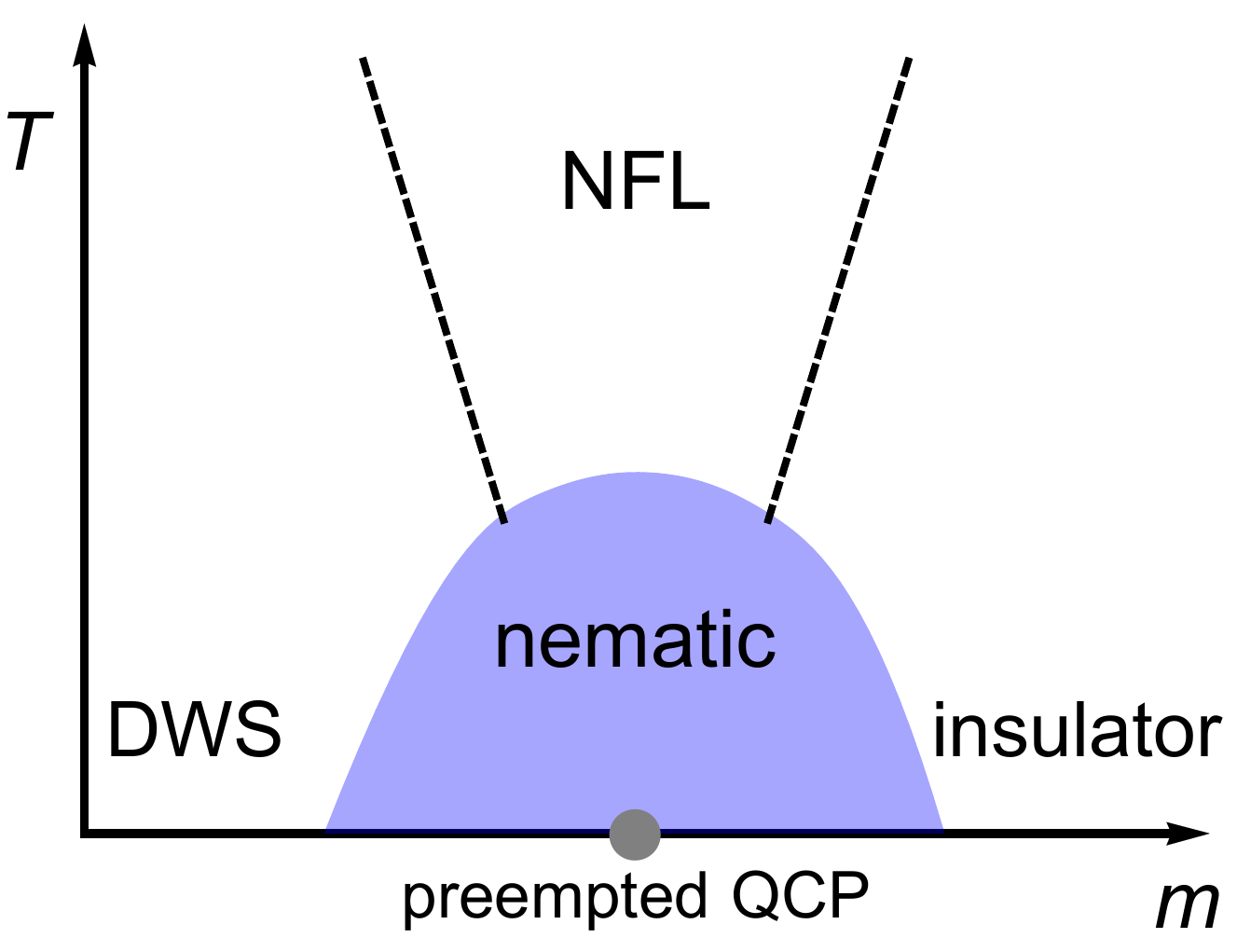}}
\caption{Schematic phase diagrams of QCPs preempted by superconductivity (a) and by nematicity (b). In (b), the QCP is preempted by nematicity in the presence of Coulomb interactions. $x$ and $m$ represent tuning parameters such as doping level and magnetic field, and $T$ refers to temperature. NFL denotes non-Fermi liquid, SC superconductivity, and DWS double-Weyl semimetals. }
\label{analogy}
\end{figure}

\textit{Introduction.}---Quantum critical phenomena are long-standing topics in condensed matter physics as universal properties and exotic physics often emerge near quantum critical points (QCPs) \cite{Subirbook,Sondhi-RMP,Herbutbook,Lohneysen2007,Stewart2001,Stewart2006}. Nonetheless, under certain circumstances, a QCP could be preempted by another symmetry-breaking phase, e.g. superconductivity as shown in \Fig{analogy}(a), such that the universal (non-Fermi-liquid) properties controlled by the putative QCP can only been measured in the critical regime {\it outside} the preempting phase. Experimental evidences of such QCP preempted by superconductivity have been reported in various systems including high-temperature superconductors (for a review, see, e.g., Refs. \cite{Taillefer2010,Shibauchi2014}). Interesting aspects of the interplay between strong fluctuations of QCP and emergent preempting phases in metallic systems with large Fermi surfaces have been extensively studied theoretically (see, e.g. Refs. \cite{Chubukov2001,Huh2008,Senthil2008,Millis2001, Sachdev2010,Metlitski2010a,Metlitski2010b,Chubukov2013,Raghu2014,Lee2015, Kivelson2015, Senthil2015,Schlief2017,Lunts2017}). However, novel features of preempted QCPs in topological semimetals remain largely unexplored.

Topological semimetals feature band-crossing points in momentum space, which are protected by their topological characters and/or crystalline symmetries \cite{Hasan2010,XLQi2011,Armitage2018,XGWan2011, GXu2011,Burkov2011,Hosur2013,nagaosanc,CFang2012,SMHuang2015b, Soluyanov2015, Ryu2016, Bradlyn2016,Ruan2015,Ruan2016}. It has been known that correlation effects in ideal topological semimetals with only discrete points at the Fermi level should be qualitatively different from the usual systems with large Fermi surfaces \cite{Shankar} because of the vanishing density of states in ideal topological semimetals. Systems hosting discrete Fermi points with either short-range interactions \cite{Herbut2006, Herbut2009, KaiSun2009, QingLiu2010, Joseph2014, Savary2014, Murray2015,Tsai2015, Roy2016} or long-range Coulomb interactions \cite{Goswami2011,Isobe2012,Isobe2013} have been extensively studied in the past decade, showing various novel behaviors such as non-Fermi liquid states \cite{Abrikosov1971, Abrikosov1974, Moon2013,Moon2018}, topological Mott insulators \cite{Herbut2014, Janssen2015, Janssen2016a, Janssen2016b}, anisotropic screening of Coulomb interactions \cite{Abrikosov1972, BJYang2014,Isobe2016, SKJian2015,HHLai2015, SXZhang2016, Moon2018b}, fermion-induced QCPs \cite{fiqcp1,fiqcp2,fiqcp3, Scherer2017a, Scherer2017b, Scherer2018}, and even emergent spacetime supersymmetry \cite{SSLee2007, KunYang2010, Grover2014, Ponte2014, SKJian2014, SKJian2016,Yao2017}.

Family of topological semimetals includes multi-Weyl semimetals hosting double-Weyl (triple-Weyl) fermions with $\pm 2$ ($\pm 3$) monopole charge of Berry curvature in momentum space, which are generalizations of Weyl fermions with monopole charge $\pm 1$ \cite{Weyl, SYXu2015,BQLv2015b,LXYang2015, BQLv2015,SYXu2015b}. Topological phase transitions between the semimetals and insulators through the annihilation of Weyl or multi-Weyl nodes with opposite chiralities are intriguing partly because there is no expected spontaneous symmetry breaking to occur and conventional Landau's theory cannot be directly applied to describe this type of QCPs. Therefore, it is interesting and urgent to explore this type of novel QCPs by asking questions such as: Does the QCP exhibit non-Fermi liquid behaviors and to what extent is the QCP stable against generic interactions?

Here we investigate the nature of putative topological phase transitions between double-Weyl semimetals and (trivial or Chern) insulators in the presence of Coulomb interactions. We focus on intriguing aspects such as possible mechanism to preempt such putative QCPs. In the presence of long-range Coulomb interactions, our renormalization-group (RG) analysis shows that the QCP is stable exhibiting non-Fermi liquid behaviors when short-range interactions that allowed by symmetry are {\it artificially} neglected. However, we find that this putative non-Fermi-liquid QCP is unstable when short-range interactions are {\it correctly} included in the RG analysis. Specifically the putative QCP is preempted by emergent nematic phases \cite{Kivelson1998} that are induced collaboratively by long-range and short-range interactions. Around the putative QCP, the long-range part of the Coulomb interaction induces strong nematic susceptibility and nematicity emerges when short-range interactions are correctly taken into account, preempting the putative QCP as shown in Fig. \ref{analogy}(b).

\begin{figure}[t]
\includegraphics[width=8.5cm]{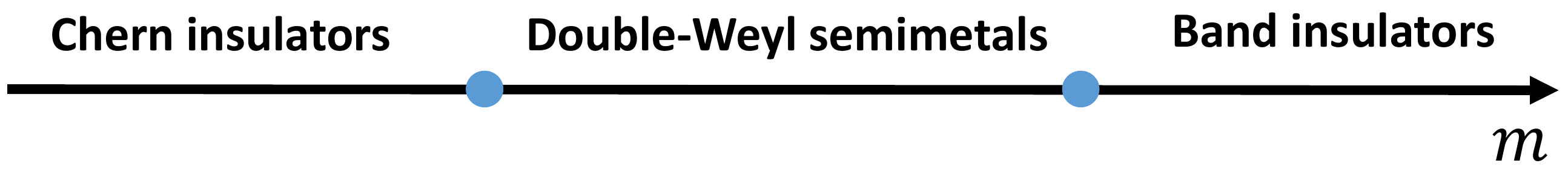}
\caption{The quantum phase diagram with varying $m$ for the model in Eq. \eqref{mlattice}.}\label{mphase}
\end{figure}

\textit{The model.}---We consider a two-band model of non-interacting fermions on cubic lattice exhibiting topological phase transitions between double-Weyl semimetals and insulators:
\bea\label{mlattice}
H_0=\sum_\vk c_{\vk}^\dagger\Big[2t_1 (\cos k_y-\cos k_x)\sigma_x+2t_2 \sin k_x\sin k_y \sigma_y\nn\\
~~+t_3(6-2\cos k_x-2\cos k_y-2\cos k_z+m)\sigma_z\Big]c_\vk,~~~
\eea
where $\sigma_i$ are Pauli matrices representing orbital degrees of freedom, $c_{\vk \alpha}^\dagger$ create spin-polarized electrons in $\alpha=1,2$ orbitals, and $t_j$ with $j=1,2,3$ denote various hopping amplitudes. We have set lattice constant to one for simplicity and we assume $t_1\!=\!t_2$ hereafter as their difference is not essential to our discussions below. The parameter $m$ can be tuned by experimental knobs such as pressure or magnetic field to access different phases, including double-Weyl semimetal (DWS), three-dimensional (3D) Chern insulator (CI), and trivial band insulator (BI). The quantum phase diagram of this non-interacting Hamiltonian as a function of $m$ is shown as Fig. \ref{mphase}. The Hamiltonian in Eq. (\ref{mlattice}) respects $C_{4h}$ symmetries apart from translational symmetries. In the DWS phase, it is the $C_4$ rotational symmetry around the $z$-axis that protects the double-Weyl fermions; the mirror symmetry ($z\rightarrow -z$) requires two double-Weyl nodes have the same energy.

As shown in Fig. \ref{mphase}, $m=0,-4$ represent the non-interacting QCPs between DWS and insulators (BI or CI). The QCPs realize quadratic band touching (QBT). Note that the QBT at the QCP is still anisotropic between $k_x/k_y$ and $k_z$ directions due to the lack of cubic symmetry. We call QBT fermions at such QCPs as critical quadratic fermions (CQF). They are critical states achieved by fine-tuning some parameter, say $m$ in the present case. Therefore, CQF is qualitatively different from the stable 3D QBT systems, such as pyrochlore iridates and $\alpha$-Tin, which are protected by $\rm{O}_h$ point-group symmetry and described by the Luttinger Hamiltonian with isotropic dispersions~\cite{Luttingerh}. We shall focus on the quantum critical point at $m=0$ below, and the same physics applies to the critical point between the DWS and CI.

The QCP of non-interacting fermions in Eq. \eqref{mlattice} is stable against weak short-range interactions because of the vanishing density of states at Fermi level (see the Supplemental Materials). However, since the density of states at the Fermi level vanishes, the Thomas-Fermi mechanism may fail to sufficiently screen Coulomb interactions. We need to carefully investigate whether Coulomb interactions are effectively screened or not in such system, especially at the putative QCP between the DWS and the insulator. As pointed out in previous works \cite{SKJian2015,HHLai2015}, deep in the double-Weyl semimetal phase the strength of long-range tail of Coulomb interactions is marginally irrelevant, rendering double-Weyl semimetal a marginal Fermi liquid. However, CQF have larger densities of states in low energy which is expected to be more susceptible to interactions than double-Weyl fermions. Therefore, it is desired to study the fate of CQF in the presence of Coulomb interactions by performing RG analysis.

\textit{RG analysis of preempted QCP.}---It is worth noting that as long as there is finite long-range Coulomb interaction, short-range interactions can be generated at low energy even when their bare values are zero. This is because the short-range four-fermion interaction can be generated at one-loop level from long-range Coulomb interaction (see the Feynman diagram in the SM). Therefore, one need to consider both long-range interactions as well as short-range interactions simultaneously at the beginning, and see how the interplay between long-range and short-range interactions affects the QCP in question.

We are ready to write down the effective field theory in the continuum including both long-range and short-range parts of the Coulomb interaction. The long-range part of the Coulomb interaction can be represented by introducing a boson field $\phi$. The (Euclidean) action at the putative QCP is then given by
\bea\label{action}
&&S\!=\!\int \frac{\md^3k \md \omega}{(2\pi)^4} \Big[ \psi^\dagger_\vk(i\omega \!+\! {\cal H}_{0\vk})\psi_{\vk} \!+\!\re{2}\phi_\vk(k_x^2+k_y^2+\eta k_z^2)\phi_{-\vk} \Big]\nn\\
&&~~~~~~~~+\int\md^4 x \Big[i e\phi\psi^\dagger\psi + g(\psi^\+\psi)^2 \Big],
\eea
where ${\cal H}_{0\vk}=t_1(k_x^2-k_y^2)\sigma_x+2t_2k_xk_y\sigma_y+t_3k_z^2\sigma_z$ represents the low-energy effective Hamiltonian of the non-interacting lattice model at the QCP ($m=0$ or $4$), and $e$ and $g$ stand for the strength of long-range Coulomb interaction and short-range interactions (there is only one independent on-site four-fermion interaction term), respectively. Note that the parameter $\eta>0$ is introduced in the kinetic term of boson fields to reflect the generic anisotropy of Coulomb potentials between the $x/y$ and $z$ directions. The hopping parameters $t_1$ and $t_2$ are in general different as the lattice system respects only the discrete $C_4$ rotational symmetry. When $t_1=t_2$, a U(1) rotational symmetry in the $xy$ plane emerges in the low-energy effective action in \Eq{action}.

We then perform RG analysis of the effective theory in \Eq{action} to derive critical behaviours of the putative QCP in the presence of Coulomb interaction. We set the scaling dimensions $[\omega]=1$, $[k_{x,y}]=z_1$ and $[k_z]=z_3$ to keep the non-interacting part invariant under RG. In general, $z_1$ and $z_3$ are different due to the anisotropy between $x/y$ and $z$-directions. We obtain $z_1,z_3$ by requiring $t_1$ and $t_3$ fixed (namely the flow equations for $t_i$ equal zero). The remaining RG equations for various parameters in the action are given by (see the SM for details):
\bea
\frac{\md e}{\md l}&=&(-\frac{z_3}{2}+\frac{1}{2}-\frac{\eta_\phi}{2})e,\label{RG1}\\
\frac{\md \eta}{\md l}&=&(2z_1-2z_3-\eta_\phi)\eta+ F_{\eta},\label{RG2}\\
\frac{\md g}{\md l}&=&(1-2z_1-z_3)g+F_1 g^2+F_2 ge^2+F_3 e^4,\label{RG3}
\eea
where $\eta_\phi$ is the anomalous dimension of the boson field $\phi$ and $F_i$ are some numerical functions derived from Feynman diagram amplitude (see SM for their definition).

When the long-range part of Coulomb interaction is not present ($e=0$), it is clear that the short-range interaction $g$ is irrelevant at the Gaussian QCP between the semimetals and insulators. When $e>0$, the system may fail to screen the long-range Coulomb interaction effectively due to the vanishing density of the states at the putative QCP. As a consequence, the long-range Coulomb interaction can render non-trivial correlation effect at the putative Gaussian QCP as we analyze below.

\begin{figure}
\includegraphics[width=5.5cm]{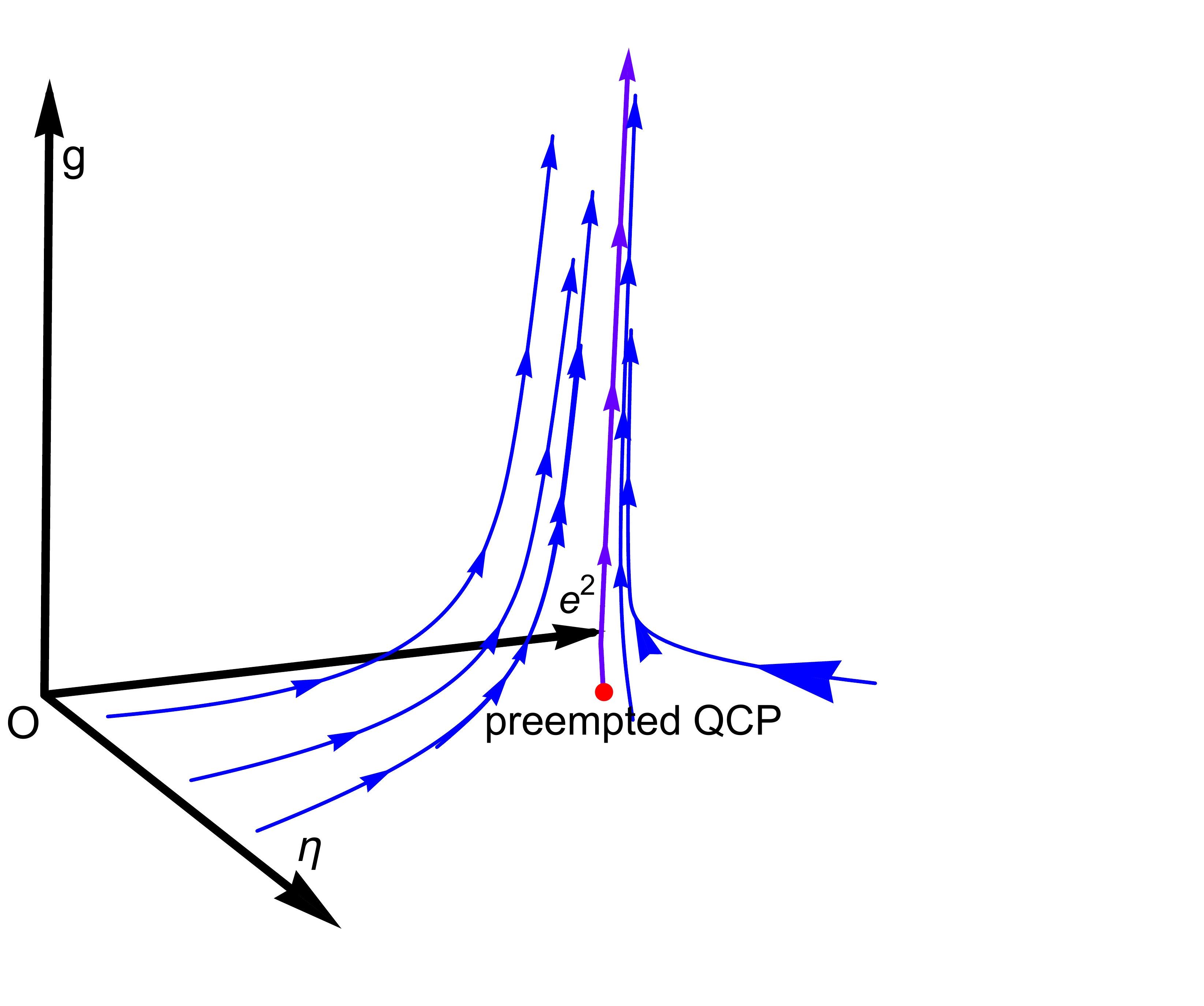}
\caption{The RG flow diagram of the critical quadratic fermions with both short-range and long-range interactions. The red point stands for the NFL fixed point when we artificially discard short-range interactions in RG analysis. It is clear all the flows lead to strong coupling of short-range interaction $g$ as long as $e>0$ which preempts the presumed QCP with the NFL fixed point. }
\label{flow}
\end{figure}

When $e>0$, it turns out that short-range interaction $g$ cannot be neglected in the RG analysis even when its bare value is zero ($g_0=0$). This is because the long-range part of the interaction can generate short-range interaction $g$ under RG flow, as clearly shown in \Eq{RG3}. However, if one \textit{artificially} restricts RG flows within the parameter space of $g=0$, one obtains an exotic QCP which corresponds to a non-Fermi-liquid fixed point characterized by anisotropic Coulomb interaction given by $e\neq 0$ and $\eta\approx 2/3$ (see the SM for details). This non-Fermi-liquid QCP, obtained by requiring $g_0=0$ and artificially neglecting the flow of $g$, is marked as the red point in the $g=0$ plane, as shown in Fig. \ref{flow}.

However, the putative non-Fermi-liquid fixed point in the $g=0$ plane is actually unstable once the short-range interaction $g$ is correctly allowed to flow under RG, as shown by the run-way trajectory in Fig. \ref{flow}. Since the short-range interaction is allowed by symmetry, its bare value $g_0$ is in general nonzero. Even when its bare value is fine tuned to zero, it is inevitably generated by the long-range part $e$ of the Coulomb interaction. Consequently, one must include both long-range and short-range interactions simultaneously when exploring the low-energy universal physics around the putative quantum phase transition. Even infinitesimal long-range Coulomb interactions are able to drive the flow of the short-range interaction to strong-coupling limit. The runaway RG flow of short-range interactions implies that certain type of symmetry breaking should occur around the putative QCP although the RG flow itself cannot tell which type of ordering actually is induced. After knowing the relevant interactions under the RG flow, one can employ the mean-field calculations to obtain the pattern of symmetry breaking. We find that the putative Gaussian QCP between the semimetals and insulators is destroyed by (even infinitesimal) Coulomb interactions and intermediate nematic phases emerge between the semimetals and insulators. In other words, the presumed QCP is preempted by nematicity.

\textit{The quantum phase diagram.}---Since the QCP is shown to be preempted by nematic ordering, a natural question is how low-energy physics near the QCP gets modified. For double-Weyl fermions near the presumed QCP, the separation of two double-Weyl nodes at $\pm \vk^\ast=(0,0,\pm\sqrt{|m|})$ is small. Before the annihilation of double-Weyl nodes, the low-energy physics of the system is captured by the interplay between long-range Coulomb and short-range interaction of the double-Weyl fermions.

\begin{figure}[t]
\includegraphics[width=5.5cm]{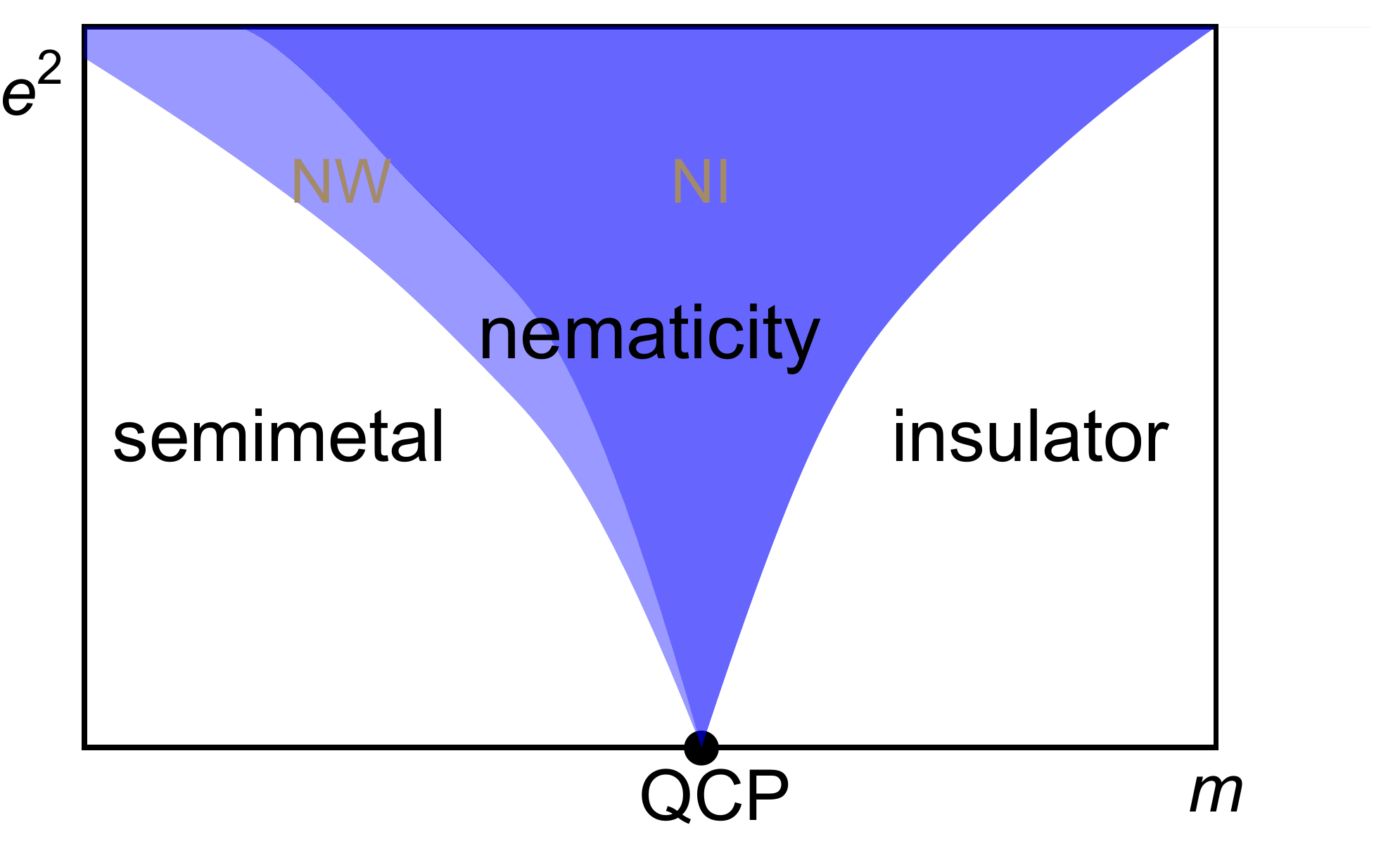}
\caption{Quantum phase diagram describing annihilation of double-Weyl fermions in the presence of Coulomb interaction. The putative QCP between (double-Weyl) semimetals and (trivial or Chern) insulators is preempted by nematic phases as long as Coulomb interaction is finite. NI and NW denote nematic insulator and nematic Weyl semimetal, respectively.}
\label{phasediag}
\end{figure}

The Hamiltonian of the double-Weyl fermion around $\vk^\ast_\pm$ in continuum can be deduced from Eq. \eqref{mlattice}. We first consider the double-Weyl fermion around $+\vk^\ast$. For $|\tilde k_z| \ll 2\sqrt{|m|}$ with $\tilde k_z=k_z-k^\ast_z$, one can obtain the following low-energy effective Hamiltonian for the double-Weyl fermion around $+\vk^\ast$: ${\cal H}_{\rm{DWF},\vk}= t(k_x^2-k_y^2)\sigma_x +2tk_xk_y\sigma_y +2\sqrt{|m|}\tilde k_z\sigma_z$, where higher order terms in $\tilde k_z$ are neglected. The cutoff of the continuous Hamiltonian for double-Weyl fermions is $\Lambda\sim \sqrt{|m|}$. The action of the double-Weyl fermions with both long-range and short-range interactions is similar to the one in Eq. \eqref{action}, except that the Hamiltonian ${\cal H}_{0\vk}$ of the CQF is replaced by ${\cal H}_{\rm{DWF}}$, namely ${\cal H}_{0\vk}\to {\cal H}_{\rm{DWF},\vk}$ in \Eq{action}. In the DWS phase, it is known that long-range Coulomb interactions are marginally irrelevant at the stable fixed point with $e\!=\!0,\eta\!=\!0$. Consequently, weak Coulomb interaction is unable to drive short-range interactions to strong coupling to destabilize DWF phase. However, when $e$ exceeds a critical value $e^*$, it can generate a relevant short-range four-fermion term that drives the system to the strong coupling and then induce a phase transition to nematic phase. Since the only scale in the system is set by $\Lambda$, one expects the critical value for Coulomb interaction scale as $e^{*2}\sim \Lambda\sim \sqrt{|m|}$ (see the SM for details). Note that this scaling analysis is consistent with the preempted QCP: $e^{*}=0$ for $m=0$. The obtained schematic quantum phase diagram is shown in \Fig{phasediag}.

\textit{Discussions and concluding remarks.}---From RG analysis, we obtained a novel picture describing the topological phase transition from 3D double-Weyl semimetals to insulators (including 3D Chern insulators). The conventional picture for this topological phase transition is simple, namely two double-Weyl nodes with opposite chiralities approach to each other and annihilate at a high-symmetry point in the Brillouin zone, rendering a fully gapped insulator after the annihilation. This picture is valid in the absence of long-range Coulomb interaction. However, when the long-range part of Coulomb interaction (even infinitesimal) is taken into account, each double-Weyl node will split into two Weyl points with the same chirality, breaking the lattice $C_4$ rotational symmetry before annihilation. Then, these split Weyl points with opposite chiralities in the nematic phase can annihilate with one another, resulting in a fully gapped insulator with nematic ordering. The physics of QCPs preempted by nematicity may be understood in the following heuristic way. If two double-Weyl fermions meet forming critical quadratic fermions, the density of states at low energy increases which is in general unfavored when relevant interactions are present and when there are other available phases with lower density of states. Indeed, by splitting each double-Weyl nodes into two Weyl nodes, the density of state is lowered such that the splitting is more favored than annihilating directly.

The preempted QCP scenario applies similarly to the presumed topological phase transition between triple-Weyl semimetals with monopole charge $\pm 3$ protected by the $C_6$ symmetry and insulators. For this case, the long-range Coulomb interaction is relevant and drives the non-interacting critical triple-Weyl fermions to a non-Fermi-liquid fixed point, which in return renders short-range interactions relevant. The run-away flow of short-range interactions leads to nematic ordering where each triple-Weyl node is split into three Weyl points breaking the $C_6$ symmetry down to $C_3$. Therefore, in the presence of long-range Coulomb interaction, the presumed QCP where two triple-Weyl fermions annihilate each other directly is preempted by nematicity.

The picture of QCPs preempted by nematicity illustrated above could be closely related to realistic materials hosting ideal multi-Weyl fermions. There are already proposals of candidate materials hosting double-Weyl fermions based on first-principle calculations including HgCr$_2$Se$_4$ \cite{GXu2011} and SrSi$_2$ \cite{SMHuang2015b}. We believe that materials realizing ideal double-Weyl or triple-Weyl semimetals might realize the preempted QCP proposed in the present work under certain circumstances. For instance, applying strain, pressure, or magnetic field to such semimetal materials should be able to tune the parameter $m$ and drive them towards insulators. One can measure quantities such as angle-dependent specific heat and angle-dependent resistivity to observe the predicted nematicity before entering symmetry-preserving insulators.

It is worth mentioning some analogies as well as distinctions between QCPs preempted by nematicity proposed in the present work and QCPs preempted by superconductivity observed in superconducting materials including high-temperature superconductors. For the latter, when approaching the preempted QCP, the instability towards superconductivity is enhanced by the strong fluctuations around the underlying non-Fermi-liquid fixed point; but the QCP itself survives under the superconducting dome although the putative non-Fermi-liquid nature of QCP is preempted due to the formation of superconductivity. However, for the former case studied here, the topological QCP itself disappears and is replaced by intermediate nematic phase which breaks some relevant symmetries. This may shed light to deeper understanding of the interplay between quantum phase transitions and strong correlations in topological states of matter \cite{Hasan2010,XLQi2011,XGWen2017}.

{\it Acknowledgement}: We thank S.-E. Han and Eun-Gook Moon for helpful discussions. This work is supported in part by the MOST of China under Grant Nos. 2016YFA0301001 and 2018YFA0305604 (H.Y.), and by the NSFC under Grant No. 11474175 (S.-X.Z., S.-K.J., H.Y.).

\begin{widetext}
\section{Supplemental Materials}
\renewcommand{\theequation}{S\arabic{equation}}
\setcounter{equation}{0}
\renewcommand{\thefigure}{S\arabic{figure}}
\setcounter{figure}{0}

\subsection{A. The mean-field analysis for short-range interactions}

We study the lattice model in the main text with only short-range interactions. In general, on-site short-range interactions in the two band model can be described as four-fermion interactions with no momentum dependence: $ (\psi^\dagger M\psi)(\psi^\dagger N\psi) $  where $ M, N $ are two by two Hermitian matrix and $ \psi\!=\!(c_{1\vk},c_{2\vk}) $.  In our specific systems, by requiring $ C_4 $ rotation symmetry protecting double-Weyl nodes and particle-hole symmetry which fix Fermi energy on the Weyl nodes, we are finally left with only four interactions $ (\psi \sigma_i\psi)^2 $, where $ \sigma_i $ is identity matrix for $ i=0 $ and Pauli matrix for $ i=1 $ to $ 3 $. Namely, only those interactions with $ M=N $ keep all necessary symmetry in our model. We further utilize the Fierz identity for two by two matrix  as
\eq{(\psi^\dagger M \psi)(\psi^\dagger N\psi)=-\re{4}(\mathrm{Tr} M\sigma_iN\sigma_j)(\psi^\dagger\sigma_i\psi)(\psi^\dagger\sigma_j\psi).}{fierz}
We can get four equations for interactions where we set $ M=N=\sigma_i $ and  find the unique solution which satisfies Fierz identity and symmetry requirements.
The relation is $ (\psi^\dagger\sigma_0\psi)^2=-(\psi^\dagger\sigma_i\psi)^2 $ for  $ i=1 $ to $ 3 $ and we finally reduce 10 terms of four-fermion interactions to one independent term. This term is just Hubbard interaction as $ 2g n_1n_2 $, where $ n_i=\psi_i^\dagger \psi_i $ is the density for $ i $th orbital. We always assume $ g>0 $ namely repulsive Hubbard interaction. And that can be justified by RG analysis, where the only stable run-away flow for (critical) double-Weyl fermion system is toward $ g\ra +\infty $.

In RG sense, the strength of such four-fermion interaction $ g $ has scaling dimension $ -1 $ in tree-level in double-Weyl fermion case and scaling dimension $ -1/2 $ in tree-level in critical quadratic Weyl fermion case and hence irrelevant at the Gaussian fixed point representing  free (critical) double-Weyl fermions. Namely, infinitesimal short-range interactions cannot drive the system to other phases, and only  short-range interactions with finite interaction strength exceeding some critical value $ g_c $ can induce phase transitions in this system.

Therefore, we apply mean-field approach to investigate ordered phases induced by short-ranged interactions. In principle, for a two-band model, there are four independent terms for possible orders as $ \langle\psi^\dagger \sigma_i\psi\rangle $ in particle-hole channel (Particle-particle channel instabilities are not favored since there is always a repulsive interaction). Amongst them, $ \langle\psi^\dagger\sigma_0\psi \rangle$ is just the shift of chemical potential and can be dropped. Similarly, $ \langle\psi^\dagger\sigma_3\psi\rangle  $ coupled to $ \psi^\dagger\sigma_3\psi $ corresponds the shift of $ m $ in the model. However, since we assume $ m $ is a controllable external parameter, the renormalization is also omitted. In sum, there are only two remaining order parameters which are responsible for nematic orders breaking $ C_4 $ rotation symmetry down to $ C_2 $.

We decouple the Hamiltonian with Hubbard interactions as
\al{H_{\textbf{mf}}=\sum_{\vect{k}} [(\cos k_y-\cos k_x+2g_1\Delta _1)\sigma_x+( \sin k_x\sin k_y+2g_2\Delta_2) \sigma_y+\notag\\(6-2\cos k_x-2\cos k_y-2\cos k_z+m)\sigma_z]-g_1 \Delta_1^2-g_2 \Del_2^2,\label{mf}}
where $ \Delta_i=\<\psi^\dagger \sigma_i\psi\> $ as two order parameters and $ g_1,g_2 $ are interaction strength which obey the constraint $ g_1+g_2=-g $ from Fierz identity. Our task is to minimize the free energy numerically for each $ m $ and $ g $ and find corresponding orders $ \Delta_{1,2} $. For simplicity, we assume hopping parameters $ t_1=t_2=t_3=1 $ in most of the calculations below.

In our model, when there is nematic order, it  always tends to develop $ \Delta_2\neq 0 $ phase while $ \Delta_1=0 $, and this feature is model dependent. It is worth noting that there are different phases corresponding to nematic orders $ \Delta_{2}\neq 0 $. When $  0<g_2\Delta_2<l_c$, the double-Weyl node split into two Weyl fermions in the $ xy $ diagonal directions forming nematic Weyl fermion phase; when $ g_2\Delta_2=l_c $, the four Weyl fermions meet with each other on $ k_z=0 $ plane forming so-called anisotropic Weyl fermions; and when  $ g_2\Delta_2>l_c $, there is fully gap in the system as a nematic insulator.

Similar with CQF, which is formed when two double-Weyl fermions overlap, we have anisotropic Weyl fermions(AWF) when $ g=g'_c $. AWF is formed when two single-Weyl fermions overlap and has linear dispersion in two directions and quadratic dispersion in the third momentum direction. AWF here serves as a critical state separating Weyl semimetal and nematic insulator phases which can also be named as critical Weyl fermions.

The mean-field phase diagram considering short-range interactions is shown as Fig. \ref{sphase}. Apparently, the original scenario for topological phase transitions accomplished by annihilating double-Weyl fermions remains unchanged when short-range interactions are small.

\begin{figure}[t]
\includegraphics[width=7cm]{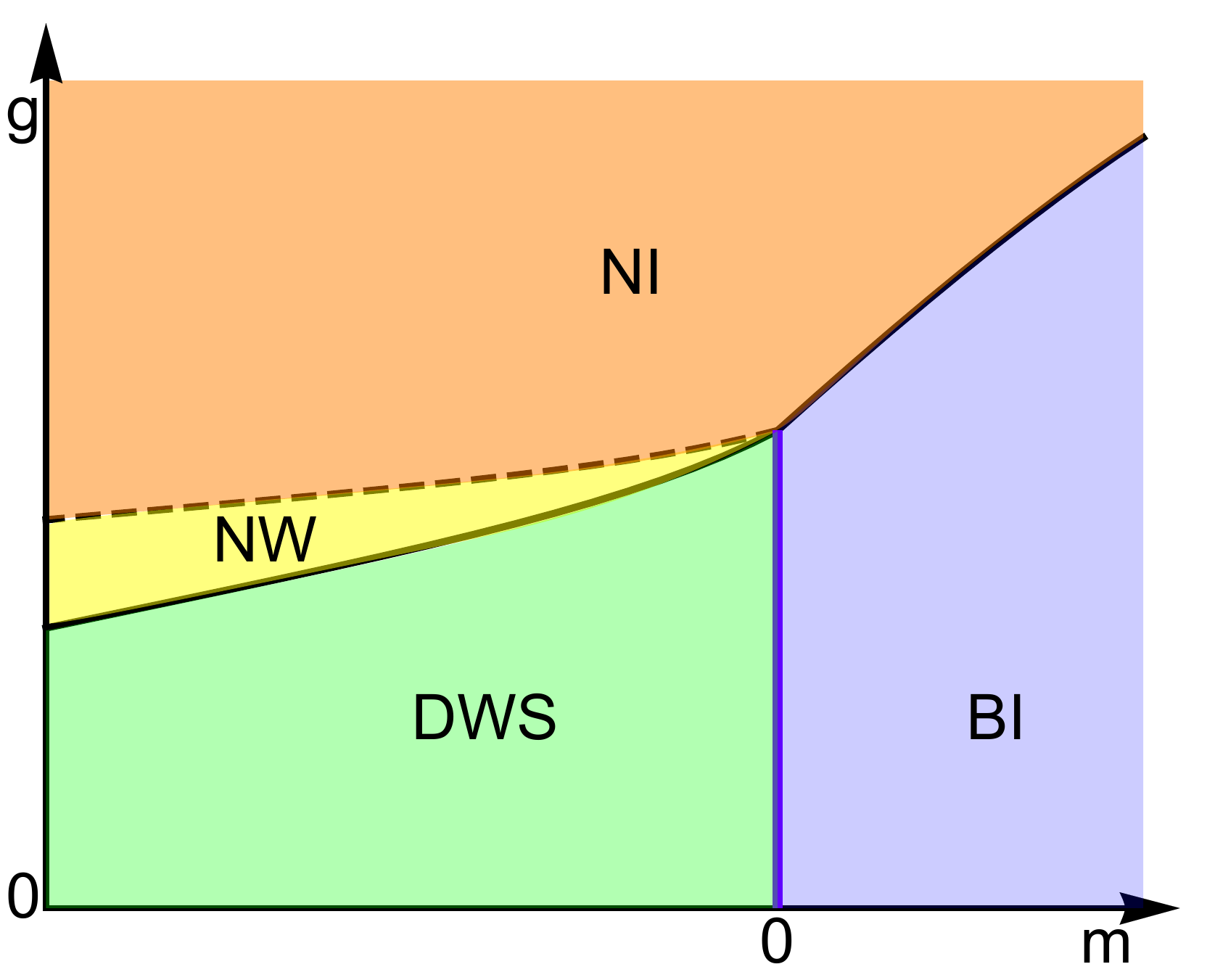}
\caption{Mean-field phase diagram with repulsive on-site interactions $ g $: BI: trivial band insulator or 3D Chern insulator. DWS: double-Weyl semimetals hosting two double-Weyl nodes. NW: nematic Weyl fermion phase. NI: nematic insulator phase. Black lines represent second order phase transitions from disorder to nematic order phase. Purple line represents topological phase transition from double-Weyl semimetals to trivial insulators whose low-energy effective theory is critical qudratic fermions(CQF). Dashed line lies at where nematic Weyl fermions annihilate as AWF. }
\label{sphase}
\end{figure}

\subsection{B. RG analysis on CQF with Coulomb interactions}
Although short-range interactions are inevitable generated from RG as we show in the next section, we here perform RG considering only Coulomb interactions to see the presume QCP and related non-Fermi liquid behaviors, which  is helpful to understand the physics when short-range interaction are considered: how NFL properties get destroyed and how the QCP is preempted.

Therefore, we carry out RG calculation on CQF systems ($ m=0 $) with Coulomb interactions alone, though short-range interactions inevitably grow, we omit them in this step. The action is captured by
\al{S=&S_\psi+S_\phi+S_e,\label{faction}\\S_\psi=&\re{(2\pi)^4}\int\md^3k \md \omega \;\psi^\dagger_\vk(-\mi\omega + H_l(m=0))\psi_{\vk},\label{psi}\\S_\phi=&\re{(2\pi)^4}\int\md^3k \md \omega \;\re{2}\phi_\vk(k_x^2+k_y^2+\eta k_z^2)\phi_{-\vk},\label{phi}\\S_e=&\int\md^4 x\;\mi e \phi\psi^\dagger\psi,\label{e}.}

\begin{figure}
\includegraphics[width=8cm]{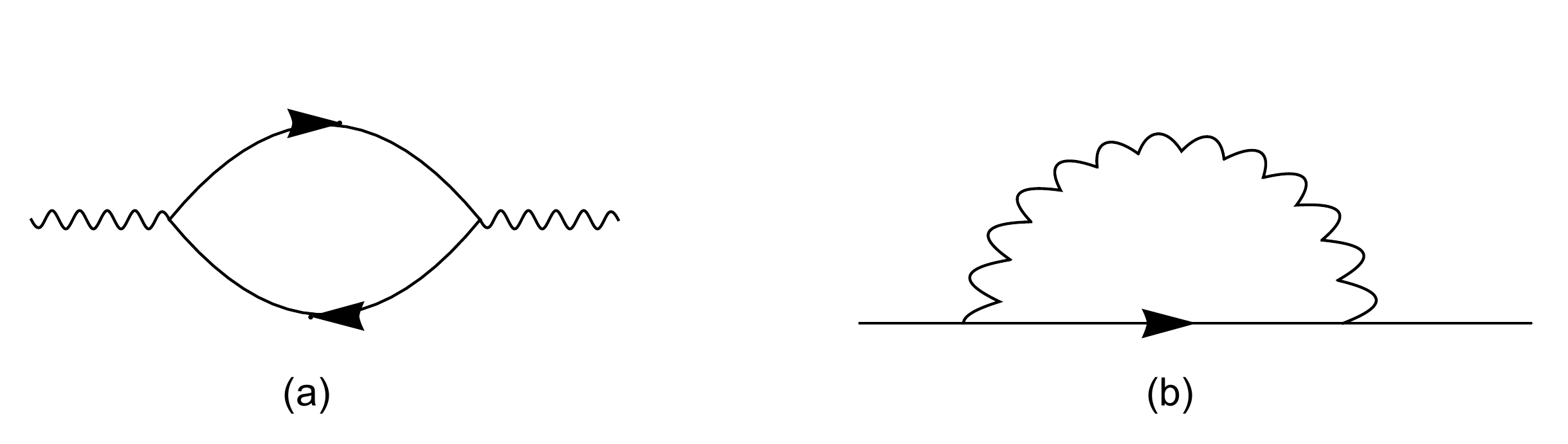}
\caption{Feynman diagrams relevant to Coulomb interactions: Solid lines stand for fermions and wavylines stand for Coulomb potential.}\label{f1}
\end{figure}

In Wilsonnian RG, integrating out the high-energy modes, will generate an effective action with lower energy cutoff and new parameters. We derive the RG equation from iteratively integrating momentum shells whose fermions are within momentum $ Q_\perp \in (Q\me^{-l},Q) $ in a infinite cylinder geometry, where $ l > 0 $ is the RG running parameter. Since the vertex correction are zero due to Ward identity, there are only two Feynman diagrams related to corrections on self-energy, see Fig. \ref{f1}. We calculate those two diagrams and compare coefficients before each term with original field theory as
\al{S=\int \psi^\dagger(\mi\omega\sigma_0+ (k_x^2-k_y^2)\delta t_1\sigma_x +2k_xk_y\delta t_1\sigma_y+k_z^2\delta t_3\sigma_z)\psi +\re{2}\phi(\eta_\phi (k_x^2+k_y^2)+\delta \eta k_z^2)\phi, }
where intergral measure is omitted and the cutoff is assumed to be unity in the calculation.

The key part in RG is the scaling dimensions. As we mentioned in the main text, though CQF disperse quadratic in three directions there are still anisotropy in three directions. Therefore, we set the scaling dimension for time-space as $ [\omega]=1 $, $ [k_{x,y}]=z_1 $ and $ [k_z]=z_3 $. We further have dimensions for other parameters as $ [\eta]=2z_1-2z_3-\eta_\phi $, $ [e]=-z_3/2+1/2-\eta_\phi/2 $, $ [t_1]=1-2z_1 $, $ [t_3]=1-2z_3 $. And we obtain $ z_1,z_3 $ by requiring $ t_1=t_3=1 $ fixed (flow equation for $ t_i $ equal zero). The remaining RG equations are
\eq{\frac{\md e}{\md l}=(-z_3/2+1/2-\eta_\phi/2)e,~~~~\frac{\md \eta}{\md l}=(2z_1-2z_3-\eta_\phi)\eta+F_\eta,}{lab}
where $F_\eta$ is from the contribution of Fig.\ref{f1}(b), by Taylor expansion on $k_z^2$.

By numerically iterating the above flow equations, we find the unique stable fixed point $ (\eta,e)\approx (0.66, 4.1) $ with finite interaction strength ($ e\neq 0$) and anisotropy for Coulomb potentials $ \eta\neq 1 $.

There is another term as $ t_s (k_x^2+k_y^2)\sigma_z $ which is also symmetry allowed in the effective Hamiltonian for low-energy fermions. We omit this term when we transform the lattice model to the effective theory for CQF. We here justify the omission of this term. The $ \beta $ function for this term is $ \md t_s/\md l=(1-2z_1)t_s+\delta t_s \approx -0.3 t_s+0.01$, where we have replaced those parameter by values on the stable fixed point. Namely, although $ t_s $ can be generated though its bare value vanishes, we can still treat it as zero safely. Because $ t_s $ is irrelevant with a negative scaling dimension and also the stable $ t_s $ is very small and we believe it has no qualitative modifications on the RG picture above.

We also mention some physical consequences here for this stable fixed point. We investigate the effect of finite $ \eta\neq 1 $ by RPA analysis.  The particle-hole polarization with propagator for CQF gives numerical results as
\eq{\Pi(q_\perp)\propto q_\perp^2, ~~~\Pi(q_z)\propto q_z^2.}{lab}when the momentum transfer is small.
The power law behaviors are the same in different directions in CQF case while there are different power laws in different directions in double-Weyl fermions. The only anisotropy in particle-hole polarization appears in the coefficients before momenta, namely we have the full polarization as
\eq{-\Pi(\vect{q})\approx a q_\perp^2+b q_z^2,}{lab}
where $ a\neq b $ representing the anisotropy in CTWF which is a weaker type of anisotropy compared to triple-Weyl fermions. Moreover, the renormalized Coulomb potential in this case behaves as
$V(\vect{q})=\re{q_\perp^2+q_z^2-\Pi(\vect{q})}\propto \re{c q_\perp^2+q_z^2}$, where $ c\neq 1 $ shows the anisotropy in Coulomb interactions. By Fourier transformations into real space, Coulomb potential behaves as
\eq{V(\vect{r})\propto \re{\sqrt{r_\perp^2+c r_z^2}}.}{ctwfrv}
The long-range behaviors of renormalized Coulomb potential together with the finite $ g $ at the non-Fermi liquid fixed point show  that Coulomb interactions receive no effective screening and actually drive the system to a non-Fermi liquid critical phase with finite interactions and the remaining anisotropy for Coulomb potential shows the difference between CQF here and 3D QBT systems given by Luttinger Hamiltonian.

In NFL states, various physical observables scale with exotic power laws. As for specific heat, consider the free CQF without Coulomb interactions, its specific heat can be deduced by densities of states near Weyl nodes $ \rho(\epsilon)\sim \epsilon^{1/2} $, which behaves as $ C\sim T^{3/2} $. When Coulomb interactions are taken into consideration, non-Fermi liquid behaviors emerge where scaling dimension $ z_1, z_3 $ get modifications from tree-level value, and specific heat in the interacting case scales as exotic power law:
\eq{C\sim T^{2z_1+z_3}\sim T^{1.82}.}{lab}

In sum, we have quantum critical point picture slightly modified by Coulomb interactions when short-range interactions are negligible (less than the critical value $ g_c $ mentioned in the last section), as illustrated in Fig. \ref{lphase}.

\begin{figure}[t]
\includegraphics[width=7cm]{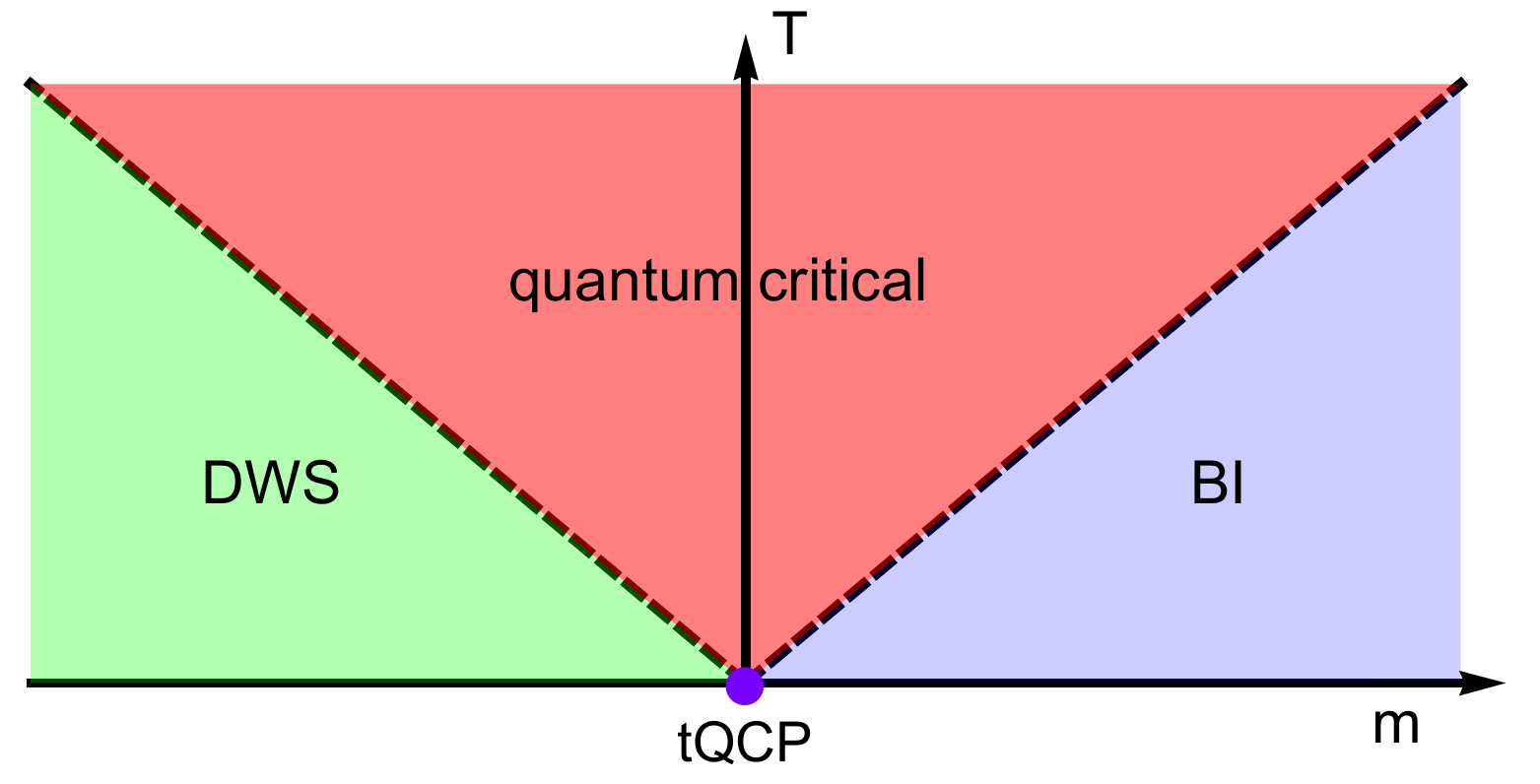}
\caption{Generic finite temperature phase diagram for quantum phase transitions in 3D systems with double-Weyl fermions: The critical quadratic Weyl fermions (CQF) emerge at the quantum critical point with $ m=0 $.  The finite-temperature crossover is described by the dashed lines and controlled by CQF at the quantum critical point. The three regimes show characteristic behaviors in physical quantities. For instance, the specific heat shows $ C\sim \me^{-|m|/T} $ in the	insulator phase, $ C\sim T^{2} $ along with exotic logarithmic corrections in the double-Weyl SM phase, and $ C\sim T^{1.82} $ in the quantum critical regime which shows non-Fermi liquid behavior.}
\label{lphase}
\end{figure}

\subsection{C. RG analysis on CQF with both Coulomb and short-range interactions}
In this section, we include both long-range interaction $ e $ as well as short-range interaction $ g $ into the full action Eq. \eqref{faction} and perform equal-footing renormalization analysis to see how the interplay between long-range and short-range interactions affect the physics picture we originally assumed.

\begin{figure}
\includegraphics[width=12cm]{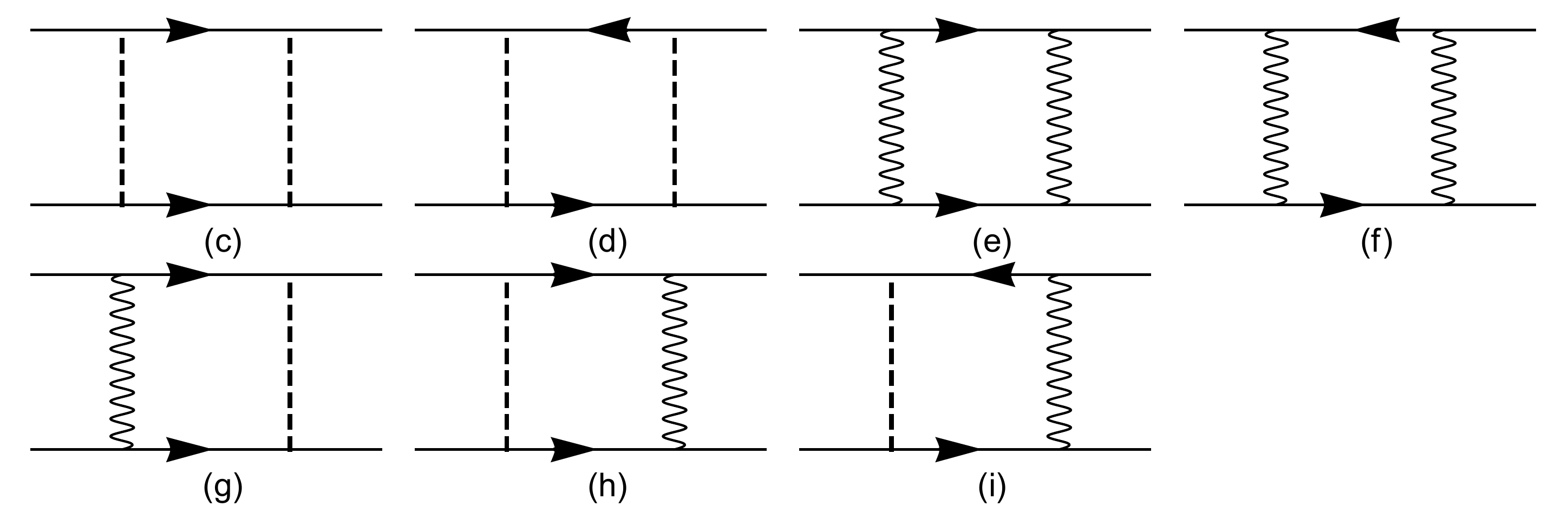}
\caption{Feynman diagrams contribute to four-fermion interactions:Solid lines stand for fermions, wavylines stand for Coulomb potential and dashed lines stand for short-range interaction. }
\label{f2}
\end{figure}

The calculation is similar with the case above except  the short-range interactions in this part. So we only focus on the renormalizations for short-range interactions in this section. There are very limit diagrams with non-vanishing amplitudes for four-fermion interactions and they are listed as Fig. \ref{f2}. Remember that we choose only one independent interaction $ g(\psi^\dagger\psi)^2 $, and once we meet other forms of interactions, we should transform them back to $ g $ using Fierz identity. The only difference compared to the last section is the inclusion of beta function for $ g $:
\eq{\frac{\md g}{\md l}=(1-2z_1-z_3)g+F_1 g^2+F_2 ge^2+F_3 e^4,}{lab}
where $F_i $ are calculated from Feynman amplitudes as Fig. \ref{f2}. (c, d contribute to $F_1$, e, f contribute to $F_3$ and g,h,i contribute to $F_2$). And the existence of (e), (f) tells us  Coulomb interactions can drive out short-range interactions even when its bare value is zero. That is the key of the break-down of the conventional picture refer to this type of topological phase transitions. The flow diagram in this case is shown in the main text.

Apart from the run-away flow of short-range interaction, we note that the strength and anisotropy of Coulomb interactions remain the same as the NFL case, and the reason is that $ g $ cannot enter the flow equation for $ e $ and $ \eta $ at one-loop level. Henceforth, though the critical point is finally preempted by nematic phase, NFL behaviors might still be accessible  in some coupling parameter regions as similar scenarios in HTS.

\subsection{D. RG near the QCP: double-Weyl fermions with both interactions}
As explained in the main text, we use an effective theory for double-Weyl fermions to investigate behaviors around but not exactly at the QCP. And we use the implicitly assumed cutoff in the action as the control parameter which tunes the separation of two double-Weyl nodes in crystal momentum space: $ \Lambda\sim\sqrt{m} $. All the above RG procedures still apply to double-Weyl fermion case in principle as long as we replace the propagator for CQF with  double-Weyl fermions. And note this time we cannot simply set cutoff $ \Lambda $ to be unity. Instead, we need to vary $ \Lambda $  to study the scaling behavior for phase boundaries around the QCP.

The first observation is the existence of critical $ e^* $. Coulomb interactions is marginally irrelevant in double-Weyl fermion case, which means infinitesimal Coulomb interactions cannot drive short-range interactions leaving the system in the double-Weyl fermion phase. However, Coulomb interaction exceeding $ e^* $ can still lead to run-away flow of on-site interaction. This picture can be directly shown from the flow diagram Fig. \ref{estar}.

\begin{figure}
\includegraphics[width=7cm]{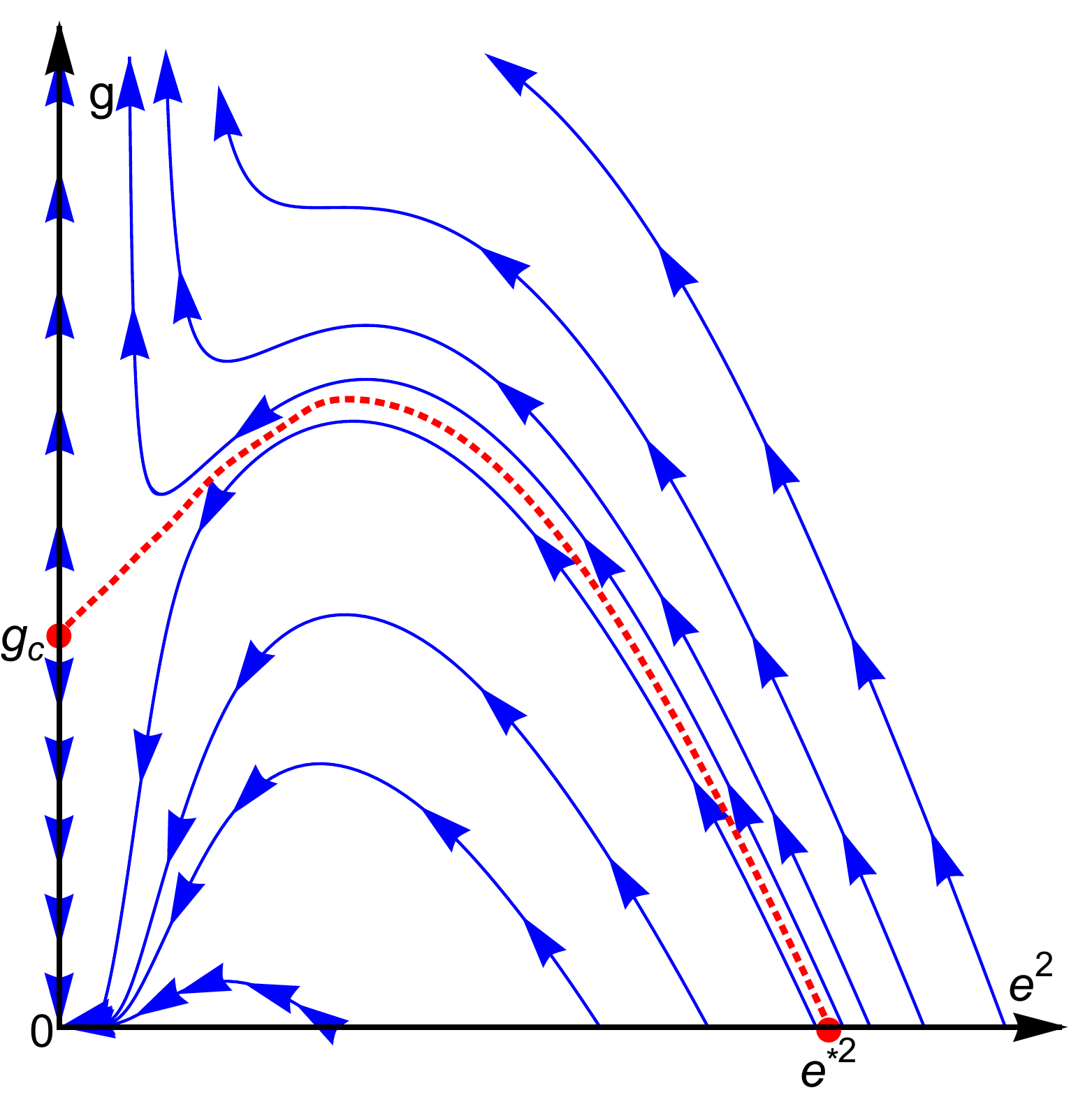}
\caption{Flow diagrams for double-Weyl fermions with both types of interactions: The red dotted line stand for the phase boundary between double-Weyl fermions and nematic phases determined by RG.}
\label{estar}
\end{figure}

There is other information in the flow diagram. Even if $ e<e^* $, Coulomb interaction is also helpful to enhance the short-range interaction $ g $. Namely, the critical value  $ g_c $ is still finite when $ 0<e<e^* $, while it is less than the mean-field critical value now. According to the red line we come to the conclusion $ 0=g_c(e\geq e^*)<g_c(0<e<e^*)<g_c(e=0)=g_c $ in a wide parameter range.

Furthermore, we explore the scaling relations between those phase boundaries and cutoff representing the separation of double-Weyl nodes. Firstly, we can show $ e^{*2}(|m|)\sim \sqrt{|m|} $. Namely, as two double-Weyl nodes leave each other, the critical $ e $ to drive out on-site interactions becomes larger. This is consistent with the CQF limit ($ m=0 $), where we can treat it as $ e^{*}=0 $ (infinitesimal Coulomb interaction is enough to drive short-range interactions). Now consider cases with finite fixed e, due to the relation $ e^{*2}=\sqrt{|m|/C'} $ ($ C' $ is just a constant), we have the critical $ |m|_c=C' e^{4} $. And the system is at nematic phase even if there is no bare on-site interaction when $ 0<|m|<|m|_c $.  When $|m|> |m|_c $, $ e $ is now less than $ e^*(|m|) $, however, based on the above observation, $ g_c(m) $ is still less than its mean-field value. Numerical results show the scaling behavior here is $ g_c\sim \sqrt{|m|}-\sqrt{|m|_c} $ when $ m $ is slightly larger than $ m_c $.

Based on all the above results, we obtain the illustrative phase diagram for the model in the main text.

\end{widetext}

\begin{thebibliography}{99}


\bibitem{Sondhi-RMP} S. L. Sondhi, S. M. Girvin, J. P. Carini, and D. Shahar, Rev. Mod. Phys. {\bf 69}, 315 (1997).
\bibitem{Subirbook} S. Sachdev, {\it Quantum Phase Transitions}, Cambridge University Press, Cambridge (2011).
\bibitem{Herbutbook} I. Herbut, {\it A Modern Approach to Critical Phenomena}, Cambridge University Press, New York (2007).

\bibitem{Lohneysen2007} H. v. Lohneysen, A. Rosch, M. Vojta, and P. Wolfle, Rev. Mod.
Phys. {\bf 79}, 1015 (2007).


\bibitem{Stewart2001} G. R. Stewart, Rev. Mod. Phys. {\bf 73}, 797 (2001). \bibitem{Stewart2006} G. R. Stewart, Rev. Mod. Phys. {\bf 78}, 743 (2006).
\bibitem{Taillefer2010}L. Taillefer,  Annu. Rev. Condens. Matter Phys. {\bf 1}, 51 (2010).
\bibitem{Shibauchi2014}T. Shibauchi, A. Carrington, and Y. Matsuda, Annu. Rev. Condens. Matter Phys. {\bf 5}, 113 (2014).

\bibitem{Chubukov2001} A. Abanov, A. V. Chubukov, and A. M. Finkelstein, Europhysics Letters {\bf54}, 488 (2001).
\bibitem{Millis2001} R. Roussev and A. Millis, Phys. Rev. B {\bf63}, 140504 (2001).
\bibitem{Huh2008} Y. Huh and S. Sachdev, Phys. Rev. B {\bf 78}, 064512 (2008).
\bibitem{Senthil2008} T. Senthil, Phys. Rev. B {\bf 78}, 035103 (2008).
\bibitem{Sachdev2010} M. A. Metlitski and S. Sachdev, New Journal of Physics {\bf12}, 105007 (2010). \bibitem{Metlitski2010a} M. A. Metlitski and S. Sachdev, Phys. Rev. B 82, 075127 (2010).
\bibitem{Metlitski2010b} M. A. Metlitski and S. Sachdev, Phys. Rev. B 82, 075128 (2010).
\bibitem{Chubukov2013} Y. Wang and A. V. Chubukov, Phys. Rev. Lett. {\bf110}, 127001 (2013). \bibitem{Raghu2014} A. L. Fitzpatrick, S. Kachru, J. Kaplan, S. Raghu, and G. Torroba, and H. Wang, arXiv:1410.6814.\bibitem{Lee2015} S. Ghamari, S.-S. Lee, and C. Kallin, Phys. Rev. B {\bf92}, 085112 (2015). \bibitem{Kivelson2015} S. Lederer, Y. Schattner, E. Berg, and S. A. Kivelson, Phys. Rev. Lett. {\bf114}, 097001 (2015). \bibitem{Senthil2015} M. A. Metlitski, D. F. Mross, S. Sachdev, and T. Senthil, \PRB~{\bf 91}, 115111 (2015).	\bibitem{Schlief2017} A. Schlief, P. Lunts, and S.-S. Lee, Phys. Rev. X {\bf 7}, 021010 (2017).
\bibitem{Lunts2017} P. Lunts, A. Schlief,  and S.-S. Lee, Phys. Rev. B {\bf 95}, 245109 (2017).


\bibitem{Hasan2010}  M. Z. Hasan and C. L. Kane, Rev. Mod. Phys. {\bf 82}, 3045 (2010) and references therein.
\bibitem{XLQi2011} X.-L. Qi and S.-C. Zhang, Rev. Mod. Phys. {\bf 83}, 1057 (2011) and references therein.	

\bibitem{Armitage2018} N.P. Armitage, E. J. Mele,  and Ashvin Vishwanath,
Rev. Mod. Phys. {\bf 90}, 15001 (2018). \bibitem{XGWan2011} X. Wan, A. M. Turner, A. Vishwanath, and S.Y. Savrasov, Phys. Rev. B {\bf 83}, 205101 (2011). \bibitem{GXu2011} G. Xu, H. Weng, Z. Wang, X. Dai, and Z. Fang, Phys. Rev. Lett. {\bf 107}, 186806 (2011). \bibitem{Burkov2011} A. A. Burkov and L. Balents, Phys. Rev. Lett.  {\bf 107}, 127205 (2011). \bibitem{Hosur2013} P. Hosur and X. Qi. Comptes Rendus Physique {\bf14}, 857 (2013). \bibitem{nagaosanc} B.-J. Yang and N. Nagaosa, Nature Communications {\bf 5 } 4898 (2014). \bibitem{CFang2012} C. Fang, M. J. Gilbert, X. Dai, and B. A. Bernevig, Phys. Rev. Lett. {\bf 108}, 266802 (2012). \bibitem{SMHuang2015b} S.-M. Huang, S.-Y. Xu, I. Belopolski, C.-C. Lee, G. Chang, B. K. Wang, N. Alidoust, M. Neupane, H. Zheng, D. Sanchez, A. Bansil, G. Bian, H. Lin, and M. Zahid Hasan, arXiv:1503.05868. \bibitem{Soluyanov2015} A. A. Soluyanov, D. Gresch, Z. Wang, Q. Wu, M. Troyer, X. Dai, and B. A. Bernevig, Nature {\bf 527}, 495 (2015). \bibitem{Ryu2016} C.-K. Chiu, J. C. Y. Teo, A. P. Schnyder, S. Ryu, Rev. Mod. Phys {\bf 88}, 035005 (2016). \bibitem{Bradlyn2016} B. Bradlyn, J. Cano, Z. Wang, M. G. Vergniory, C. Felser, R. J. Cava, and B. A. Bernevig, Science {\bf 353} 6299 (2016). \bibitem{Ruan2015} J. Ruan, S.-K. Jian, H. Yao, H. Zhang, S.-C. Zhang, and D. Xing, Nature Communications {\bf 7}, 11136 (2016). \bibitem{Ruan2016} J. Ruan, S.-K. Jian, D. Zhang, H. Yao, H. Zhang, S.-C. Zhang, and D. Xing, Phys. Rev. Lett. {\bf116}, 226801 (2016).  


\bibitem{Shankar}  R. Shankar, Rev. Mod. Phys. {\bf 66}, 129 (1994).


\bibitem{Herbut2006} I. F. Herbut, Phys. Rev. Lett. {\bf 97}, 146401 (2006). \bibitem{Herbut2009} I. F. Herbut, V. Juri$\check{\textrm{c}}$i\'c, and B. Roy, Phys. Rev. B {\bf 79}, 085116 (2009). \bibitem{KaiSun2009} K. Sun, H. Yao, E. Fradkin, and S. A. Kivelson, Phys. Rev. Lett. {\bf 103}, 046811 (2009). \bibitem{QingLiu2010} Q. Liu, H. Yao, and T. Ma, Phys. Rev. B {\bf 82}, 045102 (2010). \bibitem{Tsai2015} W.-F. Tsai, C. Fang, H. Yao, and J. Hu, New J. Phys. {\bf 17}, 055016 (2015).
\bibitem{Joseph2014} J. Maciejko and R. Nandkishore, Phys. Rev. B {\bf 90}, 035126 (2014). \bibitem{Savary2014} L. Savary, E.-G. Moon, and L. Balents, Phys. Rev. X {\bf 4}, 041027 (2014).  \bibitem{Murray2015} J. M. Murray, O. Vafek, and L. Balents, Phys. Rev. B {\bf 92}, 035137 (2015). \bibitem{Roy2016} B. Roy, P. Goswami, and V. Juricic, Phys. Rev. B {\bf95}, 201102 (2017).  
\bibitem{Goswami2011} P. Goswami and S. Chakravarty,  Phys. Rev. Lett. {\bf 107}, 196803 (2011). \bibitem{Isobe2012} H. Isobe and N. Nagaosa, Phys. Rev. B {\bf 86}, 165127 (2012). \bibitem{Isobe2013} H. Isobe and N. Nagaosa, Phys. Rev. B {\bf 87}, 205138 (2013). 
\bibitem{Abrikosov1971} A. A. Abrikosov and S. D. Beneslavskii, Sov. Phys. JETP {\bf 32}, 699 (1971).
\bibitem{Abrikosov1974} A. A. Abrikosov, Sov. Phys. JETP {\bf 39}, 709 (1974).
\bibitem{Moon2013} E.-G. Moon, C. Xu, Y. B. Kim, and L. Balents, Phys. Rev. Lett. {\bf 111}, 206401 (2013). \bibitem{Moon2018} S. Han and E.-G. Moon Phys. Rev. B 97, 241101 (2018).

\bibitem{Herbut2014} I. F. Herbut and L. Janssen, Phys. Rev. Lett. {\bf 113}, 106401 (2014). \bibitem{Janssen2015} L. Janssen and I. F. Herbut, Phys. Rev. B {\bf 92}, 045117 (2015).  \bibitem{Janssen2016a} L. Janssen and I. F. Herbut, Phys. Rev. B {\bf 93}, 165109  (2016). \bibitem{Janssen2016b} L. Janssen and I. F. Herbut, Phys. Rev. B {\bf95}, 075101 (2017).  
\bibitem{Abrikosov1972} A. A. Abrikosov, J. Low. Temp. Phys. {\bf 8}, 315 (1972).
\bibitem{BJYang2014} B.-J. Yang, E.-G. Moon, H. Isobe, and N. Nagaosa, Nat. Phys. {\bf 10}, 774 (2014). \bibitem{Isobe2016} H. Isobe, B.-J. Yang, A. Chubukov, J. Schmalian, and N. Nagaosa, Phys. Rev. Lett. {\bf116}, 076803 (2016).\bibitem{SKJian2015}S.-K. Jian and H. Yao, Phys. Rev. B {\bf 92}, 045121 (2015). \bibitem{HHLai2015} H.-H. Lai, Phys. Rev. B {\bf 91}, 235131 (2015). \bibitem{SXZhang2016} S.-X. Zhang, S.-K. Jian, and H. Yao,  Phys. Rev. B {\bf 96}, 241111 (2017). \bibitem{Moon2018b} S. Han, G. Y. Cho, and E.-G. Moon, arXiv:1804.01547.

\bibitem{fiqcp1} Z.-X. Li, Y.-F. Jiang, S.-K. Jian, and H. Yao, Nature Communications {\bf 8}, 314 (2017). \bibitem{fiqcp2} S.-K. Jian and H. Yao, Phys. Rev. B {\bf 96}, 155112 (2017). \bibitem{fiqcp3} S.-K. Jian and H. Yao, Phys. Rev. B {\bf 96}, 195162 (2017). \bibitem{Scherer2017a} M. M. Scherer and I. F. Herbut, Phys. Rev. B {\bf 94}, 205136 (2016).
\bibitem{Scherer2017b} L. Classen, I. F. Herbut, and M. M. Scherer, Phys. Rev. B {\bf 96}, 115132 (2017).
\bibitem{Scherer2018} E. Torres, L. Classen, I. F. Herbut, and M. M. Scherer, Phys. Rev. B {\bf 97}, 125137 (2018).

\bibitem{SSLee2007} S.-S. Lee, Phys. Rev. B {\bf 76}, 075103 (2007).
\bibitem{KunYang2010} Y. Yu and K. Yang, Phys. Rev. Lett. {\bf 105}, 150605 (2010).
\bibitem{Grover2014}  T. Grover, D. N. Sheng, and A. Vishwanath, Science {\bf 344}, 280 (2014).
\bibitem{Ponte2014} P. Ponte and S.-S. Lee, New J. Phys. {\bf 16}, 013044 (2014).
\bibitem{SKJian2014} S.-K. Jian, Y.-F. Jiang, and H. Yao, Phys. Rev. Lett. {\bf 114}, 237001 (2015).
\bibitem{SKJian2016} S.-K. Jian, C.-H. Lin, J. Maciejko, and H. Yao, Phys. Rev. Lett. {\bf 118}, 166802 (2017).
\bibitem{Yao2017} Z.-X. Li, A. Vaezi, C. B. Mendl, and H. Yao, arXiv:1711.04772 (to appear in Science Advances).


\bibitem{Weyl} H. Weyl, Z. Phys. {\bf 56}, 330 (1929).

\bibitem{SYXu2015}S.-Y. Xu, I. Belopolski, N. Alidoust, M. Neupane, G. Bian, C. Zhang, R. Sankar, G. Chang, Z. Yuan, C.-C. Lee, S.-M. Huang, H. Zheng, J. Ma, D. S. Sanchez, B. Wang, A. Bansil, F. Chou, P. P. Shibayev, H. Lin, S. Jia, and M. Z. Hasan, Science {\bf 349}, 613 (2015).
\bibitem{BQLv2015b}B. Q. Lv, H. M. Weng, B. B. Fu, X. P. Wang, H. Miao, J. Ma, P. Richard, X. C. Huang, L. X. Zhao, G. F. Chen, Z. Fang, X. Dai, T. Qian, and H. Ding,  Phys. Rev. X  {\bf 5}, 031013 (2015).
\bibitem{LXYang2015} L. X. Yang, Z. K. Liu, Y. Sun, H. Peng, H. F. Yang, T. Zhang, B. Zhou, Y. Zhang, Y. F. Guo, M. Rahn, D. Prabhakaran, Z. Hussain, S. K. Mo, C. Felser, B. Yan, and Y. L. Chen, Nat. Phys. {\bf 11}, 728 (2015).
\bibitem{BQLv2015} B. Q. Lv, N. Xu, H. M. Weng, J. Z. Ma, P. Richard, X. C. Huang, L. X. Zhao, G. F. Chen, C. E. Matt, F. Bisti, V. N. Strocov, J. Mesot, Z. Fang, X. Dai, T. Qian, M. Shi, and H. Ding, Nat. Phys. {\bf 11}, 724 (2015).
\bibitem{SYXu2015b}S.-Y. Xu, N. Alidoust, I. Belopolski, Z. Yuan, G. Bian, T.-R. Chang, H. Zheng, V. N. Strocov, D. S. Sanchez, G. Chang, C. Zhang, D. Mou, Y. Wu, L. Huang, C.-C. Lee, S.-M. Huang, B. Wang, A. Bansil, H.-T. Jeng, T. Neupert, A. Kaminski, H. Lin, S. Jia, and M. Zahid Hasan, Nat. Phys. {\bf11}, 748 (2015).

\bibitem{Kivelson1998} S. A. Kivelson, E. Fradkin, and V. J. Emery, Nature {\bf 393}, 550 (1998).

\bibitem{Luttingerh} J. M. Luttinger, Phys. Rev. {\bf 102}, 1030 (1956).

\bibitem{XGWen2017} X.-G. Wen, Rev. Mod. Phys. {\bf 89}, 41004 (2017).
	
\end{thebibliography}
\end{document}